\documentclass[aps,pra,floatfix,superscriptaddress,preprint, showpacs,preprint,10pt]{revtex4}

\usepackage[pdftex]{graphicx}
\usepackage{amssymb}
\usepackage{amsmath}
\usepackage{epsfig}
\usepackage{latexsym}
\usepackage{color}
\usepackage{rotating}
\usepackage{verbatim}

\def\dtwo{d^{\,2}\!}

\begin{document}

\title{Nonlinear quantum metrology using coupled nanomechanical resonators}
\author{M.~J. Woolley}
\affiliation{Department of Physics, School of Physical Sciences,
University of Queensland, St Lucia, Queensland 4072, Australia}
\author{G.~J. Milburn}
\affiliation{Department of Physics, School of Physical Sciences,
University of Queensland, St Lucia, Queensland 4072, Australia}
\author{Carlton M. Caves}
\affiliation{Department of Physics, School of Physical Sciences,
University of Queensland, St Lucia, Queensland 4072, Australia}
\affiliation{Department of Physics and Astronomy, MSC07-4220,
University of New Mexico, Albuquerque, New Mexico 87131-0001, USA}

\begin{abstract}
We consider a nanomechanical analogue of a nonlinear interferometer,
consisting of two parallel, flexural nanomechanical resonators, each
with an intrinsic Duffing nonlinearity and with a switchable
beamsplitter-like coupling between them.  We calculate the precision
with which the strength of the nonlinearity can be estimated and show
that it scales as $1/n^{3/2}$, where $n$ is the mean phonon number of
the initial state.  This result holds even in the presence of
dissipation, but assumes the ability to make measurements of the
quadrature components of the nanoresonators.
\end{abstract}


\maketitle

\section{INTRODUCTION}

High-precision measurement is an essential component of any advanced
technology, be it classical or quantum.  In the case of the emerging
quantum technologies, however, constraints on our ability to make
precise measurements are imposed by the Heisenberg uncertainty
principle.  It is often the case that the measurement objective is
simply to estimate a single parameter of the Hamiltonian of a
system~\cite{Helstrom1}. For example, in atomic clocks the objective
is to estimate the resonant frequency of a given electronic
transition~\cite{clocks}.  In the case of an optical interferometer,
the objective is to estimate an optical phase shift produced by some
mechanism of interest, which changes the relative path length.  As
the parameter varies, the dynamics of the system changes, and precise
determination of the parameter requires that the change in the
dynamics of the system be resolvable with sufficient sensitivity to
the parameter.

The precision with which the parameter can be determined depends on
the initial state of the system, the nature of the Hamiltonian
describing the system's evolution, and the measurements to be
performed on the system.  Most work on parameter estimations assumes
that system quanta are coupled independently to the parameter,
meaning that the coupling is quadratic in the field variables, which
leads to equations of motion that are linear in the field variables.
For such linear couplings, the optimal precision in a parameter
estimate scales as $1/n$, where $n$ is the number of system quanta
used in the measurement, a scaling known as the Heisenberg
limit~\cite{Lloyd1}. Achieving the Heisenberg limit with a linear
coupling requires using an entangled initial state~\cite{Lloyd2}. If
one is restricted to using product states, then a linear coupling can
only achieve a $1/n^{1/2}$ scaling, which is called the shot-noise
limit or the standard quantum limit.

It was recently shown that quantum parameter estimation with scaling
better than $1/n$ could be attained by using a coupling to the
parameter that is \emph{nonlinear\/} in field
variables~\cite{Caves1}; such super-Heisenberg scalings can be
obtained even with an initial product state~\cite{Caves2}.  The use
of product states, as opposed to entangled states, circumvents the
difficulty of creating the entangled states and also makes the scheme
considerably more robust against the deleterious effects of
decoherence.  In a related development, a measurement of a phase
shift of an optical field~\cite{Pryde1}, using an adaptive
measurement scheme that requires no entanglement, has achieved a
Heisenberg-limited sensitivity in terms of number of interactions of
photons with the phase shifter, rather than just the number of
photons.

Flexural nanomechanical resonators have an intrinsic \emph{Duffing
nonlinearity\/} due to extension on bending~\cite{Roukes1};
technology is progressing towards the point where such resonators can
be cooled to near their quantum ground state~\cite{Lehnert1}. Thus
nanoresonators might provide a system in which parameter estimation
beyond the $1/n$ Heisenberg limit could be demonstrated, in this case
measurement of the nonlinear coefficient for the Duffing
nonlinearity.  Other candidates for the first experimental
demonstration of super-Heisenberg scalings in parameter estimation
include the measurement of phase in a nonlinear optics
setting~\cite{Luis1,Beltran1,Luis2}, the measurement of a magnetic
field in atomic magnetometry \cite{Geremia1}, and the measurement of
atomic scattering strength in coupled Bose-Einstein
condensates~\cite{Lukin1,Sundaram1,Davis1}.

\begin{figure}[ht]
\includegraphics[scale=0.7]{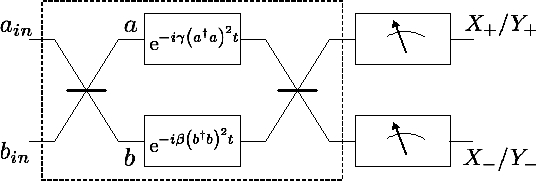}
\caption{Quantum circuit representation of nonlinear nanoresonator
interferometer. The input nanoresonator modes experience a pulsed
beamsplitter-like interaction, evolve according to a nonlinear
Hamiltonian, and the beamsplitter-like interaction is then pulsed on
again. We assume that measurements can be made of either the $X$ or
the $Y$ quadrature, of one or both output modes (denoted ``$+$'' and
``$-$''). Though not shown in the circuit, we have also considered the
effect of dissipation accompanying the nonlinear evolution.}
\label{fig0}
\end{figure}

Quite apart from any fundamental considerations, the ability to make
high-precision measurements of the Duffing nonlinearity of a
nanomechanical resonator might be of considerable practical interest.
The Duffing nonlinearity of a nanomechanical resonator is an
expression of the applied strain~\cite{Carr}.  Nonlinear
micro-electromechanical systems (MEMS) have already been used to make
highly sensitive mechanical strain sensors and accelerometers, with
applications to engineering and biomedical systems~\cite{MEMS}.
High-precision measurements of the Duffing nonlinearity would also
have implications for ultra-sensitive nanomechanical mass and force
detection~\cite{Cleland1}.

We consider two parallel, flexural nanomechanical resonators, each
with an intrinsic Duffing nonlinearity and with a switchable,
electrostatically-actuated beamsplitter-like coupling between them.
The measurement proceeds as follows: one nanoresonator is excited
into a large-amplitude coherent state, the beamsplitter interaction
is pulsed on so that the coherent-state excitation is split equally
between the two resonators, the nanoresonators evolve independently
under the nonlinearities (and standard linear dissipation), the
beamsplitter interaction is pulsed in the same way again, and a
homodyne measurement of the nanoresonator quadratures is performed.
As depicted in Figure~\ref{fig0}, this scheme effectively realizes a
nonlinear interferometer~\cite{Yamamoto1}.  We have calculated the
precision with which such a scheme can estimate the nonlinear
coefficient of one nanoresonator and demonstrated that the precision
scales as $1/n^{3/2}$, where $n$ is the mean phonon number of the
initial coherent state.

\section{SYSTEM PROPERTIES AND HAMILTONIAN}

Each nanoresonator can be regarded as a thin bar of length $l$ and
lateral width $a$.  We are interested in the fundamental mode of
vibration of each nanoresonator in the lateral direction.  Each
fundamental mode is described by a position-momentum pair,
$x_i$-$p_i$, where $i=a,b$ labels the resonators.  With a
time-dependent capacitive coupling dependent on the displacements
from the equilibrium positions, $C(x_a,x_b)$, and Duffing
nonlinearities characterized by coefficients $\chi_i$, $i=a,b$, the
system can be described classically by a Hamiltonian
\begin{equation}\label{eq:Hcl}
H_{cl} =
\frac{1}{2}m\omega^2x^2_a + \frac{p^2_a}{2m} + \frac{1}{2}m\omega^2x^2_b + \frac{p^2_b}{2m} +
P(t) \frac{1}{2}C(x_a,x_b)V^2_0 + \frac{1}{4}\chi_a m\omega^2x^4_a +
\frac{1}{4}\chi_b m\omega^2 x^4_b\;,
\end{equation}
where $P(t)$ specifies the coupling voltage pulses.

In current experiments, the nonlinear coefficient $\chi$ of a
nanoresonator is estimated~\cite{Roukes1} by measuring the critical
amplitude $a_c$ at which the forced oscillations become
bistable~\cite{Nayfeh1} and the quality factor $Q$ of the oscillator
and then using the relation $\chi=2\sqrt{3}/9a^2_{c}Q$.  Achievable
values of these parameters are $a_c=0.7\,$nm and $Q=20\,000$, giving
a nonlinear coefficient $\chi=4\times10^{13}\,\hbox{m}^{-2}$. We use
this value of $\chi$ as a typical value in the following.

We assume that the nanoresonators are capacitively coupled to nearby
bias conducting surfaces in such a way that for small displacements,
the capacitance can be expanded as
\begin{equation}
C(x_a,x_b)=C_0\!\left(1+\frac{fx_a^2+fx_b^2+2x_ax_b}{d^2}+\cdots\right)\;.
\end{equation}
Here $C_0$ is the capacitance when the oscillators are at their
equilibrium positions.  The capacitive coupling must be balanced so
that there is no net force on the resonators when the coupling is
switched on (i.e., no linear terms in the expansion).  This leaves
the quadratic terms as the dominant effect of the coupling.  In the
quadratic terms, $d\simeq100\,$nm is a characteristic lateral
separation between the resonators and the other conducting surfaces
and $f$ is a factor of order unity.  Both $d$ and $f$ depend on the
specific design of the capacitive coupling.  Provided
\begin{equation} \label{eq:kappa}
C_0V_0^2/2m\omega d^2\equiv\kappa\ll\omega\;,
\end{equation}
we can neglect the renormalization of the resonator frequencies
during the pulsing of the capacitive coupling, retaining only the
coupling between the resonators.  With these assumptions, the
capacitive term in the Hamiltonian~(\ref{eq:Hcl}) can be replaced by
$P(t)C_0V^2_0x_ax_b/d^2$, which gives rise to the desired
beamsplitter coupling.  The parameter $\kappa$, introduced in
Eq.~(\ref{eq:kappa}), characterizes the strength of the beamsplitter
coupling.

Now we quantize by introducing the operators
\begin{equation}
\hat{x}_a = ( a + a^\dagger ) \sqrt{\frac{\hbar}{2m\omega}}
\;,\qquad
\hat{p}_a = -i( a - a^\dagger ) \sqrt{ \frac{\hbar m \omega}{2}}\;,
\end{equation}
satisfying the usual commutation relations, and similarly for
nanoresonator $b$. Transforming to an interaction picture and using
the rotating-wave approximation, we find
\begin{equation}
H = \hbar \gamma(a^\dagger a)^2 + \hbar \beta(b^\dagger b)^2 + \hbar\kappa P(t)( a^\dagger b + ab^\dagger)\;,
\label{Hint}
\end{equation}
where
\begin{equation}
\gamma \equiv \frac{3}{4}\omega \chi_a (\Delta x)^2 \;,\qquad\beta \equiv \frac{3}{4}\omega \chi_b (\Delta x)^2\;,
\end{equation}
with $\Delta x = \sqrt{\hbar/2m\omega}$ being the half-width of the
ground-state wave function.  The use of the rotating-wave
approximation requires that the time scale $\delta t$ over which the
coupling is varied in $P(t)$ be significantly longer than the periods
of the nanomechanical resonators.  Since $\delta t\simeq\kappa^{-1}$,
this is the requirement, already introduced in Eq.~(\ref{eq:kappa}),
that $\kappa\ll\omega$.

At low temperatures, nanoresonator damping is thought to be mostly
due to coupling to a bath of two-level systems, but this mechanism is
not fully understood. Therefore we do not try to model this
mechanism, but rather treat dissipation using a quantum optics master
equation (with a zero-temperature bath), with the expectation that
this generic model of damping provides a reasonable account of the
effect of dissipation on parameter estimation.  Then the evolution of
the density matrix describing the state of the two nanoresonators is
given by
\begin{equation}
\dot{\rho}(t) =
-\frac{i}{\hbar}[ H, \rho] + \frac{\Gamma_a}{2}( 2a\rho a^\dagger - a^\dagger a \rho - \rho a^\dagger a)
+ \frac{\Gamma_b}{2}( 2b\rho b^\dagger - b^\dagger b \rho - \rho b^\dagger b)\;. \label{me}
\end{equation}

Experimentally reasonable values for the system properties are
$l=2\,\mu\hbox{m}$, $a = 40\,$nm, $m=10^{-17}\,$kg, $\omega = 2\pi
\times15\,\hbox{MHz}=9.4\times 10^{7}\,$rad/s, $d=120\,$nm,
$Q=20\,000$, $\chi=4\times10^{13}\,\hbox{m}^{-2}$, $C_0 = 10\,$aF,
and $V_0=1\,$V. These correspond to $\Delta x=240\,$fm, $\Gamma_i=
4\,700\,\hbox{s}^{-1}$,
$\gamma,\beta=1.6\times10^{-4}\,\hbox{s}^{-1}$ (we use
$10^{-4}\,\hbox{s}^{-1}$ as a typical value in the following), and
$\kappa=3.7 \times 10^5\,\hbox{s}^{-1}$.  The quantity
$1/2\chi(\Delta x)^2=5\times 10^{11}$ is roughly the number of
phonons required to make the quartic nonlinearity as large as the
harmonic potential; it corresponds to an oscillation amplitude
$1/\sqrt\chi=200\,$nm.

In Sec.~\ref{sec:precision} we analyze estimation of the nonlinear
coefficient $\gamma$ of oscillator~$a$, assuming that oscillator~$b$
has no nonlinearity ($\beta=0$).  Other operating conditions are
possible and yield similar results, but the figures in the remainder
of the paper refer to the $\beta=0$ case.  We consider a fiducial
evolution time $t=10^{-3}\,$s, so that $\Gamma_i t=4.7$, meaning that
the effects of dissipation are large, but not overwhelming, and we
consider a fiducial initial phonon number $n=10^7$, so that the
nonlinear phase shift $n\gamma t$ is about $1\,$rad.  We investigate
values within about an order of magnitude of these fiducial values.
Notice that a phonon number $n=10^7$ corresponds to an oscillation
amplitude $\Delta x\sqrt{2n}=1\,$nm.  This amplitude is close to the
value we assumed for $a_c$, not by accident, but because the two
oscillation amplitudes quantify, one for free oscillations and one
for forced oscillations, the same measure of the relative strengths
of the nonlinearity and the damping.

Our assumptions about the switchable beamsplitter coupling require
that $\kappa\ll\omega$, so that we can make the rotating-wave
approximation, and that $t\gg\kappa^{-1}=2.7\,\mu\hbox{s}$, so that
we can regard the beamsplitter pulses as essentially instantaneous.
Both of these inequalities are well satisfied by the above values.

\section{SYSTEM EVOLUTION IN THE Q REPRESENTATION}

Now suppose that nanoresonator $a$ is excited into a coherent state
with amplitude $\alpha_0$, assumed real. The (product) state of the
two nanoresonators is then given, in the $Q$
representation~\cite{Milburn3}, by $Q(\alpha, \alpha^*; t=0) = e^{-|
\alpha - \alpha_0|^2}/\pi$ and $Q(\beta , \beta^*; t=0) = e^{ -|
\beta|^2}/\pi$.  If we pulse on the beamsplitter interaction for a
time $\delta t = \pi/4\kappa$, the state of the system is a product
of two equal-amplitude coherent states; in other words, we have a
``balanced'' beamsplitter.  The pulse time $\delta t$ is sufficiently
short that the effect of nonlinearities and dissipation are
negligible.  We then have
\begin{equation}
Q(\alpha, \alpha^*; t=\delta t) = \frac{1}{\pi}\,e^{-| \alpha - \alpha_0/\sqrt{2}|^2}\;,\qquad
Q(\beta , \beta^*; t=\delta t) =  \frac{1}{\pi}\,e^{-| \beta - \alpha_0/\sqrt{2}|^2}\;.\label{afterBS}
\end{equation}
Setting $P(t)= 0 $ in Eq.~(\ref{Hint}) for the time between the
beamsplitter pulses, the master equation~(\ref{me}) can be converted
into a Fokker-Planck equation for the $Q$ function and solved to
give~\cite{Milburn1}
\begin{eqnarray}
Q( \alpha , \alpha^* ;t)
& = & \frac{e^{-\left|\alpha \right|^2}}{\pi}
\sum^{\infty}_{p,q=0} \frac{1}{p\,!\,q!} \left(\frac{\alpha^* \alpha_0}{\sqrt{2}}\right)^p
\left(\frac{\alpha\alpha^*_0}{\sqrt{2}}\right)^q
f_a(t)^{(p+q)/2}\exp\!\left[-|\alpha_0|^2 \frac{ f_a(t) + i\delta_a } { 2(1+i\delta_a) } \right]\;,
\label{Qalpha} \\
Q( \beta , \beta^* ;t)
& = & \frac{e^{-\left|\beta \right|^2}}{\pi}
\sum^{\infty}_{p,q=0} \frac{1}{p\,!\,q!} \left(\frac{\beta^* \alpha_0}{\sqrt{2}}\right)^p
\left(\frac{\beta \alpha^*_0}{\sqrt{2}}\right)^q
f_b(t)^{(p+q)/2}\exp\!\left[-\left|\alpha_0 \right|^2 \frac{ f_b(t) + i\delta_b } { 2(1+i\delta_b) }\right]\;,
\label{Qbeta}
\end{eqnarray}
where
\begin{eqnarray}
\delta_a = 2\gamma( p - q)/ \Gamma_a\;, &\quad& \delta_b = 2\beta( p - q)/ \Gamma_b\;, \\
f_a(t) = \exp[ -\Gamma_a t - 2i\gamma t( p - q)]\;, &\quad&
f_b\left( t \right) = \exp[ -\Gamma_b t - 2i\beta t( p - q)]\;.
\end{eqnarray}

The beamsplitter interaction is pulsed on again for time $\delta t$,
giving output quadratures
\begin{equation}
X_\pm = \frac{a + a^\dagger \pm b \pm b^\dagger}{\sqrt{2}}\;,\qquad
Y_\pm = -\frac{i(a - a^\dagger \pm b \mp b^\dagger)}{\sqrt{2}}\;.
\end{equation}
We can calculate the first and second moments of these quadratures
using Eqs.~(\ref{Qalpha}) and~(\ref{Qbeta}),
\begin{eqnarray}
\left\langle X_\pm \right\rangle & = &
\sqrt{2}\int \dtwo\alpha\,\mathcal{R}e( \alpha ) Q( \alpha, \alpha^* ; t )
\pm \sqrt{2}\int \dtwo\beta\,\mathcal{R}e( \beta ) Q( \beta, \beta^* ; t )\;, \\
\left\langle X^2_\pm \right\rangle & = &-1 +
2 \int \dtwo\alpha\,\left[ \mathcal{R}e ( \alpha ) \right]^2
Q( \alpha , \alpha^*; t ) +
2 \int \dtwo\beta\,\left[ \mathcal{R}e ( \beta ) \right]^2
Q( \beta , \beta^* ; t ) \nonumber \\
&\mbox{}&{}\quad \pm 4
\int \dtwo\alpha\,\mathcal{R}e( \alpha )Q( \alpha , \alpha^*; t )
\int \dtwo\beta\,\mathcal{R}e( \beta )Q( \beta , \beta^*; t )\;.
\end{eqnarray}
Corresponding moments of the conjugate quadratures $Y_\pm$ are given
by the same expressions with the replacement $\mathcal{R}e(\alpha)
\rightarrow \mathcal{I}m(\alpha)$.

The evaluation of these moments reduces to the calculation of two
integrals,
\begin{eqnarray}
&&\sqrt{2} \int \dtwo \alpha\,\alpha\,Q( \alpha , \alpha^*; t) =
\sqrt{n} e^{-(\Gamma_a t +nC_2)/2} e^{i(\gamma t + nD_2/2)}\;,\\
&&2 \int\dtwo\alpha\,[\mathcal{C}( \alpha )]^2 Q(\alpha , \alpha^*; t) =
1 + \frac{n}{2}e^{-\Gamma_a t} \pm \frac{n}{2} e^{-\Gamma_a t -nC_4/2}\cos( 4\gamma t + nD_4/2)\;,
\end{eqnarray}
where $n = \alpha^2_0 $, $\mathcal{C}(\alpha)=\mathcal{R}e(\alpha)$
for the upper sign and $\mathcal{C}(\alpha)=\mathcal{I}m(\alpha)$ for
the lower sign, and
\begin{eqnarray}
C_r=C_r( \gamma, \Gamma_a , t) & = &
\frac{1}{1 + (\Gamma_a / r\gamma)^2}
\left[ 1 - e^{-\Gamma_a t}\cos r\gamma t - \frac{\Gamma_a}{r\gamma}e^{-\Gamma_a t}\sin r\gamma t \right]\;, \\
D_r=D_r( \gamma, \Gamma_a , t) & = &
\frac{1}{1 + (\Gamma_a / r\gamma)^2}
\left[ \frac{\Gamma_a}{r\gamma} + e^{-\Gamma_a t}\sin r\gamma t -
\frac{\Gamma_a}{r\gamma} e^{-\Gamma_a t}\cos r\gamma t \right]\;.
\end{eqnarray}

\begin{figure}[ht]
\begin{center}
\includegraphics[scale=0.7]{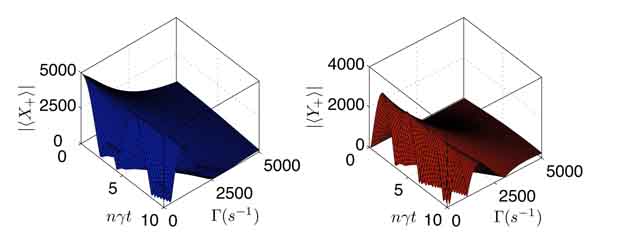}
\caption{Magnitude of expectation values of output quadratures $X_+$
and $Y_+$ as functions of the nonlinearity $\gamma$ (expressed in
terms of the nonlinear phase shift $n\gamma t$) and nanoresonator
damping $\Gamma$($=\Gamma_a = \Gamma_b$), for the choices $n=10^7$,
$\beta = 0$, and $t=10^{-3}\,$s. The rapidly oscillating fringes with
respect to $\gamma$ give rise to the enhanced sensitivity of this
parameter estimation scheme.  Damping leads to a shift in the
location of the fringes and to a decay of the quadrature
expectations.} \label{fig5}
\end{center}
\end{figure}

\begin{figure}[ht]
\begin{center}
\includegraphics[scale=0.2]{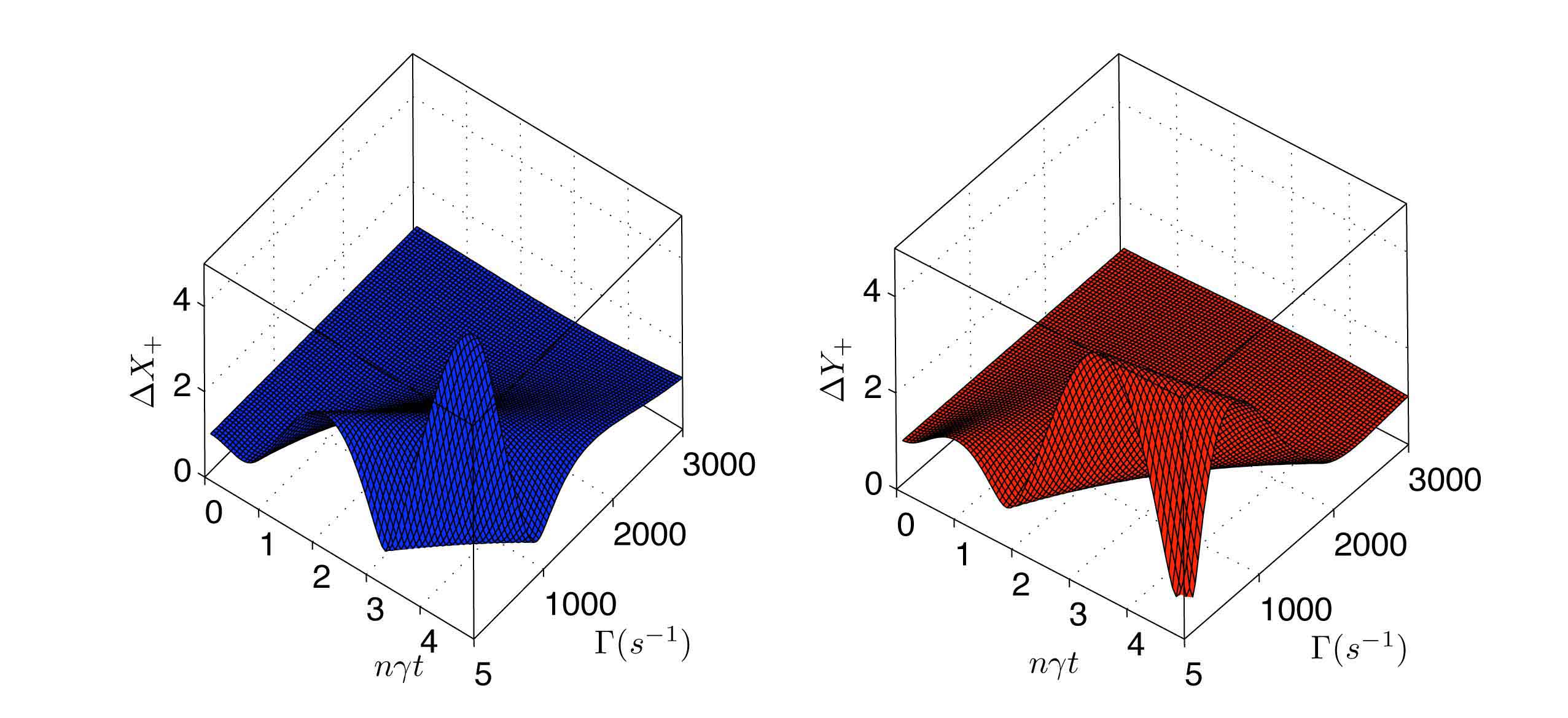}
\caption{Uncertainty in the output quadratures $X_+$ and $Y_+$ as a
function of the nonlinearity $\gamma $ (expressed in terms of the
nonlinear phase shift $n\gamma t$) and nanoresonator damping
$\Gamma$($=\Gamma_a = \Gamma_b$), for the choices $n=10^7$, $\beta =
0$, and $t=10^{-3}\,$s.  Squeezing and anti-squeezing are observed
for low damping, corresponding to shearing of the contours of the $Q$
function in phase space.  Dissipation suppresses this effect, and
coherent-state variances, corresponding to decay to the vacuum, are
seen at high damping rates.} \label{fig1}
\end{center}
\end{figure}

The general results are complicated and thus are not quoted here. In
the case of no damping ($\Gamma_a=0=\Gamma_b$), we can show that
\begin{eqnarray}
\langle X_\pm\rangle & = &
\sqrt{n}e^{-n(1-\cos 2\gamma t)/2}
\cos\!\left(\gamma t + \frac{n}{2} \sin 2\gamma t \right)\pm
\sqrt{n}e^{-n(1 - \cos 2\beta t)/2}\cos\!\left(\beta t + \frac{n}{2}\sin 2\beta t \right)\;, \\
\langle Y_\pm\rangle & = &
\sqrt{n}e^{-n(1-\cos 2\gamma t)/2}
\sin\!\left(\gamma t + \frac{n}{2} \sin 2\gamma t \right)\pm
\sqrt{n}e^{-n(1 - \cos 2\beta t)/2}\sin\!\left(\beta t + \frac{n}{2}\sin 2\beta t \right)\;,
\end{eqnarray}
and
\begin{eqnarray}
\langle X^2_\pm\rangle & = &
1 + n +
\frac{n}{2}e^{-n(1-\cos4\gamma t)/2}
\cos\!\left(4\gamma t + \frac{n}{2}\sin 4\gamma t\right) +
\frac{n}{2}e^{-n(1-\cos4\beta t)/2}
\cos\!\left(4\beta t + \frac{n}{2}\sin 4\beta t\right) \nonumber \\
&{}& \quad\pm ne^{-n(2-\cos 2\gamma t-\cos 2\beta t)/2}
\cos\!\left( \gamma t - \beta t + \frac{n}{2}\sin 2\gamma t - \frac{n}{2}\sin 2\beta t \right) \nonumber \\
&{}& \quad\pm ne^{-n( 2 - \cos 2\gamma t -\cos 2\beta t)/2}
\cos\!\left( \gamma t + \beta t + \frac{n}{2}\sin 2\gamma t + \frac{n}{2}\sin 2\beta t \right) \;, \\
\langle Y^2_\pm\rangle & = &
1 + n -
\frac{n}{2}e^{-n(1-\cos4\gamma t)/2}
\cos\!\left(4\gamma t + \frac{n}{2}\sin 4\gamma t\right) -
\frac{n}{2}e^{-n(1-\cos4\beta t)/2}
\cos\!\left(4\beta t + \frac{n}{2}\sin 4\beta t\right) \nonumber \\
&{}& \quad\pm ne^{-n(2-\cos 2\gamma t-\cos 2\beta t)/2}
\cos\!\left( \gamma t - \beta t + \frac{n}{2}\sin 2\gamma t - \frac{n}{2}\sin 2\beta t \right) \nonumber \\
&{}& \quad\mp ne^{-n( 2 - \cos 2\gamma t -\cos 2\beta t)/2}
\cos\!\left( \gamma t + \beta t + \frac{n}{2}\sin 2\gamma t + \frac{n}{2}\sin 2\beta t \right)\;.
\end{eqnarray}
Note that these results could have been calculated directly in the
Heisenberg picture with initial coherent states in each mode. Making,
in addition, the \emph{short-time approximation}, $n(\gamma
t)^2,n(\beta t)^2\ll 1$, which still allows nonlinear phase shifts
$n\gamma t$ and $n\beta t$ much larger than unity, we find the same
expectation values as for the analogous classical nonlinear
interferometer:
\begin{equation}
\left\langle X_\pm \right\rangle \rightarrow
\sqrt{n} \cos n\gamma t \pm \sqrt{n} \cos n\beta t \;,\qquad
\left\langle Y_\pm \right\rangle \rightarrow
\sqrt{n} \sin n\gamma t \pm \sqrt{n} \sin n\beta t \;.
\end{equation}

We can define a regime of strong damping by the conditions
\begin{equation} \label{eq:strongdamp}
\frac{\Gamma_a}{\gamma},\frac{\Gamma_b}{\beta}\gg\sqrt n\;,\qquad
\gamma t,\frac{\Gamma_a t}{n},\frac{\Gamma_b t}{n}\ll 1\;.
\end{equation}
The conditions on the evolution time are of little consequence
because long before they are violated, the oscillators will have
damped to the ground state.  Notice that these conditions allow the
case of most interest to us, i.e., $\Gamma_b=\Gamma_a\simeq\gamma n$,
with $n\gamma t\simeq\Gamma_a t$ allowed to be considerably larger
than unity.  In the strong damping regime, we have
\begin{eqnarray}
\langle X_\pm \rangle & = &
\sqrt{n}e^{-\Gamma_at/2}\cos\!\left[ \frac{n\gamma}{\Gamma_a}( 1 - e^{-\Gamma_a t}) \right] \pm
\sqrt{n}e^{-\Gamma_bt/2}\cos\!\left[ \frac{n\beta}{\Gamma_b}( 1 - e^{-\Gamma_b t}) \right] \;, \\
\langle Y_\pm \rangle & = &
\sqrt{n}e^{-\Gamma_at/2}\sin\!\left[ \frac{n\gamma}{\Gamma_a}( 1 - e^{-\Gamma_a t}) \right] \pm
\sqrt{n}e^{-\Gamma_bt/2}\sin\!\left[ \frac{n\beta}{\Gamma_b}( 1 - e^{-\Gamma_b t} ) \right]\;.
\end{eqnarray}
In addition, in this regime $\langle X_\pm^2\rangle=1+\langle
X_\pm\rangle^2$ and $\langle Y_\pm^2\rangle=1+\langle
Y_\pm\rangle^2$, which means that the quadrature uncertainties are at
the coherent-state level, i.e., $\Delta X_\pm=\Delta Y_\pm=1$.

The dependence of the expectation values of the ``$+$'' quadrature
amplitudes on the nonlinearity and damping rate is shown in
Figure~\ref{fig5}; similar behavior is displayed by the ``$-$''
quadratures.  The nonlinearity gives rise to rapidly oscillating
fringes, and this is the key, as we see in the next section, to the
enhanced sensitivity of the nonlinear interferometer.  Dissipation
leads to a reduction in the expectation values and also to a
reduction in the fringe frequency.  Figure~\ref{fig1} shows the
uncertainties in $X_+$ and $Y_+$, also as functions of the
nonlinearity and damping rate.  In the case of low damping, as the
nonlinear phase shift increases, the variances oscillate,
corresponding to the shearing apart and partial recurrence of the
contours of the $Q$ function in phase space. Squeezing and
anti-squeezing are apparent in the variance
oscillations~\cite{Milburn4}, and the squeezing can be quite
substantial. Damping suppresses the shearing---and, hence the
squeezing and anti-squeezing---and suppresses the quantum
interference effects that give rise to partial revivals.  Strong
damping leads to a decay of the expectation values and to quadrature
variances that take on the coherent-state value. At reasonable
damping rates, we cannot derive any benefit from the squeezing,
though neither are we adversely affected by the anti-squeezing of the
quadratures.  The deleterious effect of damping is almost entirely a
consequence of reducing the signal carried by the expectation values,
not of changing the variances.

\section{PRECISION OF PARAMETER ESTIMATE}
\label{sec:precision}

To analyze the precision of estimating the strength of the Duffing
nonlinearity, we specialize in this section to the case where
oscillator~$b$ has no nonlinearity ($\beta=0$).  We are thus
estimating the nonlinear coefficient $\gamma$ of oscillator~$a$.
Other operating conditions are possible, but we focus on this one as
a representative possibility in this section.

We phrase our results in terms of the precision in estimating the
related dimensionless parameter $\gamma t$, with $t$ regarded as
fixed.  The uncertainty in an estimate of $\gamma t$ based on
multiple measurements of a quantity $Z$---in our case, $Z$ is one of
the output quadratures---can be calculated from
\begin{equation}
\delta( \gamma t ) = t\delta\gamma =
t\frac{ \Delta Z }{|d\langle Z \rangle / d \gamma |}=
\frac{ \Delta Z }{|d\langle Z \rangle / d( \gamma t )|}\;,
\end{equation}
where $\Delta Z$ is the uncertainty in $Z$.  In the case of no
damping and again making the short-time approximation, the quadrature
variances all take on coherent-state values, i.e., $ \Delta X_\pm,
\Delta Y_\pm \rightarrow 1 $.  The precision of the estimate of
$\gamma t$ thus becomes
\begin{equation}
\delta_{X_\pm}( \gamma t) = \frac{1}{n^{3/2}|\sin n\gamma t|}\;,\qquad
\delta_{Y_\pm}( \gamma t) = \frac{1}{n^{3/2}|\cos n\gamma t|}\;.
\end{equation}
These sensitivities oscillate with the fringes produced by the
nonlinear phase shift $n\gamma t$, but they all have the same basic
scaling of $1/n^{3/2}$ with phonon number.  This scaling beats the
$1/n$ scaling achievable with a linear Hamiltonian and is consistent
with the general result~\cite{Caves2} for nonlinear Hamiltonians and
initial product states.  The factor of $n$ enhancement compared with
the standard quantum limit for linear Hamiltonians is a consequence
of the rapidly oscillating fringes in the expectation values of the
output quadratures.

\begin{figure}[ht]
\begin{center}
\includegraphics[scale=0.175]{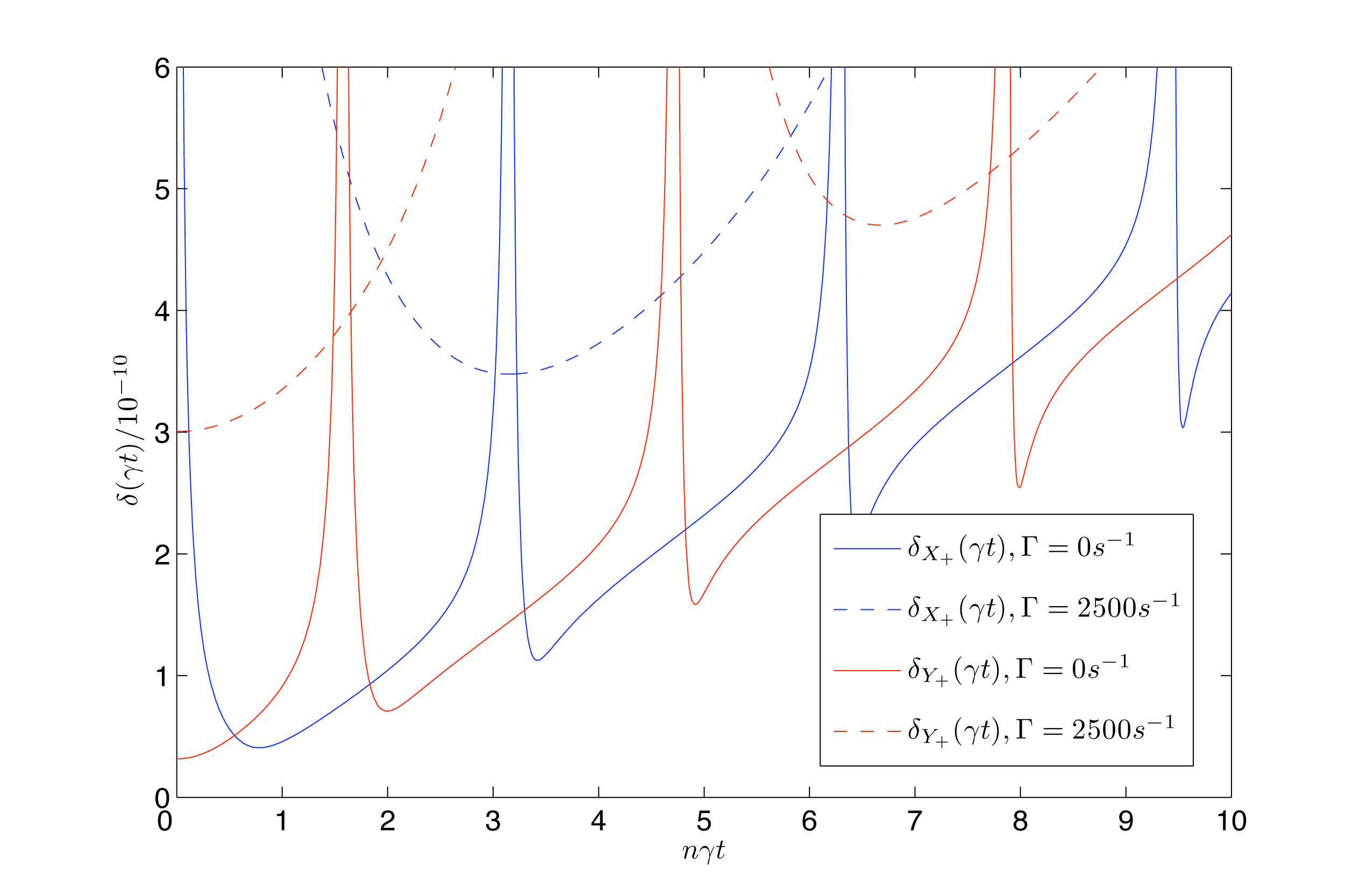}
\caption{Precision $\delta(\gamma t)$ for measurements of the $X_+$
and $Y_+$ quadratures as a function of the nonlinearity $\gamma$,
expressed as the nonlinear phase shift $n\gamma t$, for the choices
$n = 10^7$, $\beta = 0$, $t=10^{-3}\,$s, and $\Gamma_a = \Gamma_b =
\Gamma$.  Zero damping and moderate damping cases are shown for each
quadrature.  For zero damping, fringe boundaries are located at
$n\gamma t = m\pi /2$, with the fringes based on measurements of
conjugate quadratures displaced by $\pi /2 $.  Dissipation leads to
an overall reduction in sensitivity, and the fringes become more
widely spaced.} \label{fig2}
\end{center}
\end{figure}

\begin{figure}[ht]
\begin{center}
\includegraphics[scale=0.2]{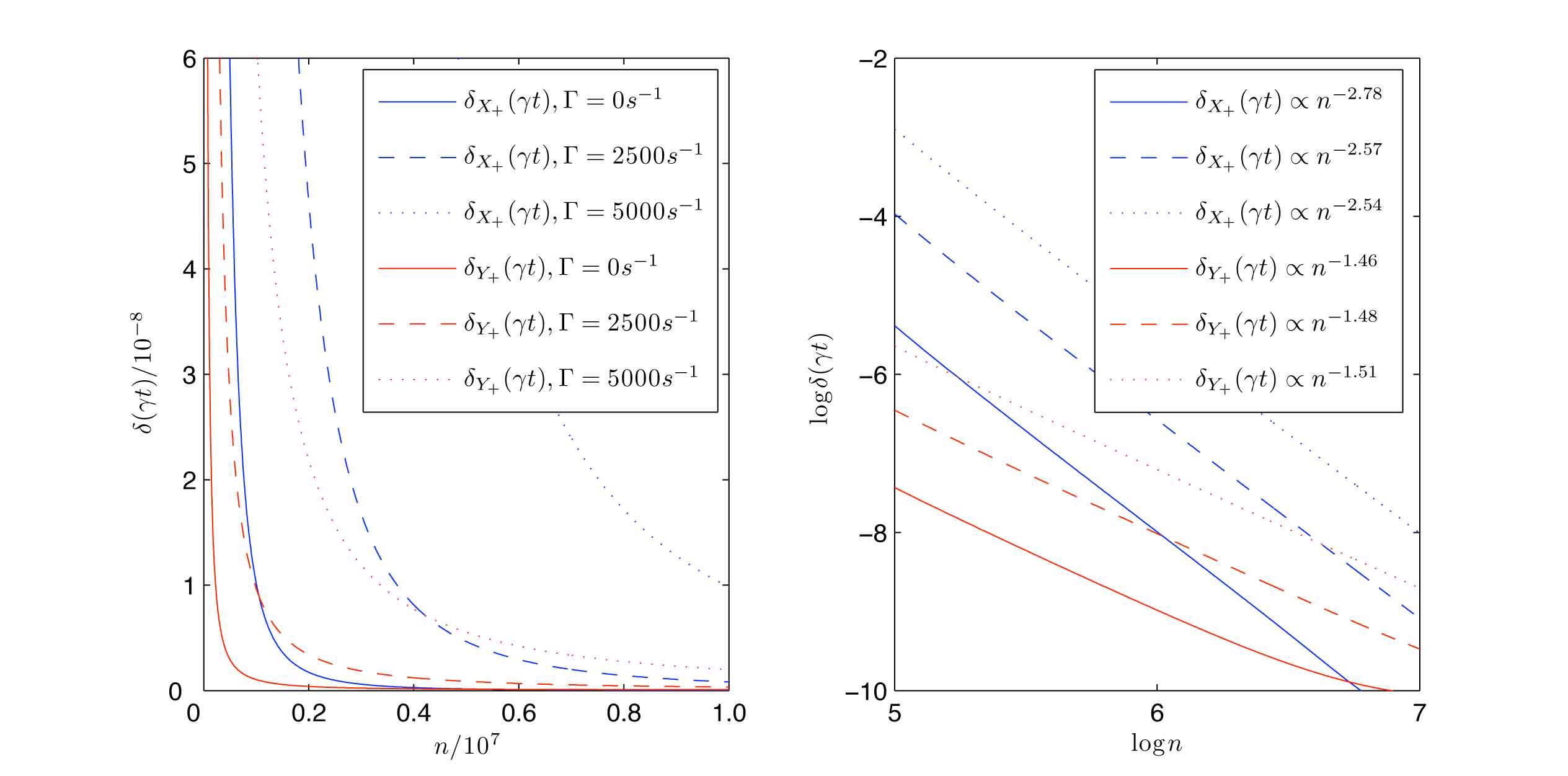}
\caption{Precision $\delta(\gamma t)$ for measurements of the $X_+$
and $Y_+$ quadratures as a function of mean phonon number $n$ in the
initial coherent state, for the choices  $\gamma =
10^{-4}\,\hbox{s}^{-1}$, $\beta = 0$, $t=10^{-3}\,$s, $\Gamma_a =
\Gamma_b = \Gamma$.  Plots for three values of the damping constant
$\Gamma$ are shown for each quadrature.  These plots correspond to
the regime $n\gamma t < 1 $.  From the log-log plots, we see that
$\delta_{X_+} \propto n^{-5/2} $ and $\delta_{Y_+} \propto n^{-3/2}$.
The extra $n^{-1}$ factor for measurement of the $X_+$ quadrature is
due to the precision improving as one moves away from the very poor
sensitivity near the central fringe boundary.} \label{fig3}
\end{center}
\end{figure}

{From} an experimental perspective, the strong damping
regime~(\ref{eq:strongdamp}) is most relevant.  In this regime, the
quadrature variances have coherent-state values, and the derivatives
of the expectation values lead to sensitivities
\begin{equation}
\delta_{X_\pm} (\gamma t) =
\frac{\Gamma_a t\,e^{\Gamma_a t/2}}{
n^{3/2}( 1 -  e^{-\Gamma_a t})
|\sin[n\gamma(1 - e^{-\Gamma_a t})/\Gamma_a]|}\;,\qquad
\label{dXdgammat}
\delta_{Y_\pm} (\gamma t) =
\frac{\Gamma_a t\,e^{\Gamma_a t/2}}{
n^{3/2}( 1 -  e^{-\Gamma_a t})
|\cos[n\gamma(1 - e^{-\Gamma_a t})/\Gamma_a]|}\;.
\end{equation}
The improved $1/n^{3/2}$ sensitivity scaling survives in the presence
of dissipation, but the absolute sensitivity is degraded, and the
fringes become more widely separated. For feasible damping rates, the
sensitivity is worsened by less than an order of magnitude, but if
the damping is further increased, the sensitivity diverges,
reflecting the absence of signal in the quadrature expectation
values.

Figure~\ref{fig2} shows the measurement precision for measurements of
the $X_+$ and $Y_+$ quadratures as a function of the nonlinearity
$\gamma$ and for two values of the damping rate
$\Gamma=\Gamma_a=\Gamma_b$.  Fringe boundaries are located at
$n\gamma t = m\pi /2 $; those based on measurement of the $X_+$ and
$Y_+$ quadratures are displaced by $\pi /2$.  As the damping rate
increases, the overall sensitivity worsens, and the fringes become
more widely spaced.  These effects can be traced back to the
reduced-amplitude and reduced-frequency oscillations of the
quadrature expectations as a function of the nonlinear phase shift.

The scaling of the measurement precision as a function of $n$ is
plotted in Figure~\ref{fig3}.  Here $n$ is chosen so that $n\gamma t
< 1 $.  The precision associated with measurement of the $Y_+$
quadrature is then near its optimal value, away from its first fringe
boundary at $n\gamma t=\pi/2$, whereas the precision associated with
measurement of the $X_+$ quadrature decreases rapidly as it falls
from the very poor sensitivity near its central fringe boundary at
$n\gamma t = 0$.  From the log-log plot, we can calculate that
$\delta_{X_+} \propto n^{-5/2} $ and $\delta_{Y_+} \propto n^{-3/2}$,
though the extra $n^{-1} $ in the $n^{-5/2}$ scaling is due to the
sensitivity falling from the central fringe boundary, and the true
scaling of the optimal sensitivity achievable is $n^{-3/2}$. The
scaling behavior is maintained in the presence of feasible levels of
dissipation, although there is a marked deterioration in sensitivity.

In practice, the initial state of the excited nanoresonator would be
better described by a displaced thermal state.  Provided that the
thermal width of this state is small compared with the amplitude of
the displacement, a condition that would be well satisfied by an
excitation at the level we are considering, the primary effect of an
initial thermal distribution would be simply to increase the output
quadrature variances, leading to a reduction in sensitivity, but not
to change in the scaling behavior.  In the strong damping regime,
contact with a finite temperature bath would result in thermal rather
than coherent state variances in the output quadratures, again
leading to a reduction in sensitivity, but not to a change in the
scaling with $n$.  Note that the quantum optics master equation for
the nonlinearities we consider and with a finite temperature bath has
been solved analytically using the $Q$
representation~\cite{Milburn2}.  We conclude that the measurement
scheme we describe is reasonably robust to increases in temperature.
A more severe difficulty lies in performing quantum-limited homodyne
detection of the output quadratures, though it is conceivable that
such measurements could be performed using coupled microwave
cavities~\cite{Lehnert2}.

\section{CONCLUSIONS}

We have calculated the precision with which the nonlinearity of a
nanomechanical resonator can be estimated, using a nanomechanical
analogue of a nonlinear interferometer.  For an input coherent state,
the precision scales as $1/n^{3/2}$, a scaling beyond that achievable
with a linear coupling even when entangled input states are employed.
This scaling behavior is maintained in the presence of dissipation,
which we modeled using a quantum optics master equation, and it is
expected that this scheme is reasonably robust to increases in
temperature.  Quantum-limited homodyne detection of the nanoresonator
quadratures is, however, a very challenging experimental task.

An alternative scheme would use a single nonlinear nanomechanical
resonator coupled to the field in a superconducting microwave cavity,
which would act as the second ``arm'' of an interferometer.  The
nonlinearity would then only be in the mechanical arm of the
interferometer.  The initial state would be excited by driving the
cavity, and read-out would be performed by quantum-limited homodyne
detection of the cavity output.  A beamsplitter-like interaction
between the cavity and nanoresonator could be realized by driving the
cavity on its red sideband~\cite{Woolley1}.  This beamsplitter
coupling would be continuous, so the analytical results obtained here
are not directly applicable.  Investigating the achievable
sensitivity of this cavity-nanoresonator scheme is the subject of
ongoing work.

\acknowledgments We thank A.~C. Doherty for useful discussions. This
work was supported in part by the Australian Research Council and by
the US Office of Naval Research Grant No.~N00014-07-07-1-0304.

\end{document}